\documentclass[12pt]{article}
\usepackage{bm}
\usepackage{amsmath}
\usepackage{amssymb}
\usepackage{float}
\usepackage{url}

\renewcommand{\thepage}{}
\makeatletter
\@addtoreset{equation}{section}
\renewcommand{\theequation}{\thesection.\@arabic\c@equation}
\makeatother
\renewcommand{\thefootnote}{\fnsymbol{footnote}}
\begin{document}
\begin{titlepage}
\title{
\vspace*{-4ex}
\hfill
\begin{minipage}{3.5cm}
\end{minipage}\\
\bf Comments on Observables for Identity-Based Marginal
 Solutions in Berkovits' Superstring Field Theory
\vspace{0.5em}
}

\author{Isao {\sc Kishimoto}$^{1}$
\ \ and\ \  
Tomohiko {\sc Takahashi}$^{2}$
\\
\vspace{0.5ex}\\
$^1${\it Faculty of Education, Niigata University,}\\
{\it Niigata 950-2181, Japan}\\
$^2${\it Department of Physics, Nara Women's University,}\\
{\it Nara 630-8506, Japan}}
\date{April, 2014}
\maketitle
%

\begin{abstract}
\normalsize
We construct an analytic solution for tachyon condensation 
around identity-based marginal solutions 
in Berkovits' WZW-like open superstring field theory.
Using this, which is a kind of wedge-based solution,
the gauge invariant overlaps for the identity-based marginal solutions
can be calculated analytically.
This is a straightforward extension of a method in bosonic string field
theory, which has been elaborated by the authors, to superstring.
We also comment on a gauge equivalence relation
between the tachyon vacuum solution and 
its marginally deformed one. From this viewpoint, 
we can find the vacuum energy of the identity-based
marginal solutions to be zero, which agrees with the previous result
as a consequence of $\xi$ zero mode counting.
\end{abstract}

\end{titlepage}

\renewcommand{\thepage}{\arabic{page}}
\renewcommand{\thefootnote}{\arabic{footnote}}
\setcounter{page}{1}
\setcounter{footnote}{0}
%
\section{Introduction}

Identity-based solutions corresponding to marginal deformations
\cite{Kishimoto:2005bs,Kishimoto:2005wv}
were constructed in Berkovits' Wess-Zumino-Witten (WZW) like open superstring
field theory \cite{Berkovits:1995ab, Berkovits:2000zj}. 
They are characterized by a simple construction even for
marginal currents with singular operator product expansions, namely
for general current algebra. 
Since the contact term divergence does not appear, 
the solutions are easily handled compared with others
\cite{Kiermaier:2007ki} based on wedge states.
Especially in the WZW-like theory, the
vacuum energy of the identify-based marginal solutions can be found to be
exactly zero, because the $\xi$ zero mode is not contained in a correlation
function in the calculation of the energy \cite{Kishimoto:2005bs,Kishimoto:2005wv}. 
As the vacuum energy was computed,
a remaining problem is to evaluate gauge invariant overlaps of the
identity-based marginal solutions in the superstring field theory.

In bosonic open string field theory, identity-based marginal
solutions 
can be represented as a difference of wedge-based solutions plus an
integration of a deformed BRST exact state \cite{Kishimoto:2013sra}.
This expression has made it possible to evaluate the gauge
invariant overlaps for the identity-based solutions in the bosonic
theory. In this calculation, the key ingredient is the ``$K'Bc$ algebra'' by
which an analytic tachyon vacuum solution is constructed in the
marginally deformed background \cite{Inatomi:2012nv}.
This provides a useful technique in string field theory
and, in fact, it is applied recently to construct a new simple
analytic solution by Maccaferri \cite{Maccaferri:2014cpa}.
Maccaferri's solution is  gauge equivalent
to the identity-based marginal solution but it is wedge-based, and so
its physical observables are analytically calculable.

The main purpose of this paper is to present an analytic calculation of
observables for the identity-based marginal
solutions in the superstring field theory.
To do this,
we will fully use the tachyon vacuum solution given by Erler
\cite{Erler:2013wda} and
a supersymmetric extension of the techniques used in the bosonic case.

In this paper, we will prove the following relation among the Erler solution
$\Phi_T^{\rm E}$, the tachyon vacuum solution $\Phi_T$ in marginally deformed
background, and the identity-based marginal solution $\Phi_J$:
\begin{align}
 e^{-\Phi_J}\hat{Q}e^{\Phi_J}=e^{-\Phi_T^{\rm E}}\hat{Q}e^{\Phi_T^{\rm E}}
-e^{-\Phi_T}\hat{Q}_{\Phi_J}e^{\Phi_T}+\int_0^1
 \hat{Q}_{\tilde{\Phi}_T(t)}\Lambda_t\,dt,
\label{idge}
\end{align}
where $\hat{Q}_{\Phi}$ is a BRST operator around a string field $\Phi$ and
$\tilde{\Phi}_T(t)$ is a classical solution interpolating from
$\Phi_T^{\rm E}$ to $\Phi_T$, and $\Lambda_t$ is a state within the
small Hilbert space.
Thanks to this relation, we will be able to calculate the gauge
invariant overlaps\footnote{
See \cite{Michishita:2004rx}, for example, in the framework of Berkovits' WZW-like superstring field theory.
}
for $\Phi_J$.
Here, we should note that the relation (\ref{idge}) implies
a gauge equivalence relation between $\Phi_T^{\rm E}$ and $\log(e^{\Phi_J}e^{\Phi_T})$. 
Namely, there exist some group elements $h$ and $g$ such that
$\hat{Q}h=0$, $\hat{\eta} g=0$ and $e^{\Phi_J}
e^{\Phi_T}=h\,e^{\Phi_T^{\rm E}}g$.
From this gauge equivalence,
denoting the action with a BRST operator $\hat{Q}$ in the NS sector as
$S[\Phi;\hat{Q}]$, we find that
\begin{align}
S[\Phi_J;\hat{Q}]&= S[\Phi_T^{\rm E};\hat{Q}] -S[\Phi_T;\hat{Q}_{\Phi_J}].
\label{veid}
\end{align}
$-S[\Phi_T^{\rm E};\hat{Q}]$ is the vacuum energy given by Erler and
$S[\Phi_T;\hat{Q}_{\Phi_J}]$ will be calculated later.
Accordingly, in addition to the gauge invariant overlaps, the vacuum energy of the
identity-based marginal solutions can be evaluated analytically.

This paper is organized as follows. In \S \ref{sec:K'Bc}, we briefly illustrate
the identity-based marginal solutions in the WZW-like theory and the
theory expanded around the solutions, which corresponds to the theory in
a marginally deformed background. Then we find a version of the extended $KBc$
algebra in the expanded theory and construct a tachyon vacuum solution
in a marginally deformed background as a variant of the Erler
solution. For the solution, we find a homotopy operator and evaluate
physical observables, the vacuum energy and the gauge invariant overlap, analytically.
In \S \ref{sec:analytic}, we prove the gauge equivalence relation mentioned
above and we calculate analytically the vacuum energy and the gauge invariant overlaps of
the identity-based marginal solutions.
In \S \ref{sec:Rem}, we give some concluding remarks.
In Appendix \ref{sec:ge}, we comment on gauge equivalence relations.

\section{Tachyon vacuum solution in the theory around the identity-based
 marginal solution
\label{sec:K'Bc}}

\subsection{Identity-based marginal solutions}

In \cite{Kishimoto:2005bs}, we have a type of
identity-based marginal solutions
in Berkovits' WZW-like superstring field theory:
\begin{align}
&\Phi_J=\tilde{V}^a_L(F_a)I,
&\tilde{V}_L^a(f)\equiv \int_{C_{\rm L}}\frac{dz}{2\pi
 i}f(z)\frac{1}{\sqrt{2}}c\gamma^{-1}\psi^a(z),
\label{marginalsol}
\end{align}
where $F_a(z)$ is some function such as $F_a(-1/z)=z^2F_a(z)$,
$C_{\rm L}$ denotes a half unit circle: $|z|=1$, ${\rm Re}\,z\ge 0$
and $I$ is the identity state.
We use a notation: $\gamma^{-1}(z)=e^{-\phi}\xi(z)$ as in
\cite{Erler:2013wda},
where $\beta,\gamma$-ghosts are expressed by 
$\xi,\eta,\phi$, such as $\beta= e^{-\phi}\partial \xi,
\gamma=\eta \,e^{\phi}$.
As for $\psi^a$ in $\Phi_J$, we suppose that
${\bm J}^a(z,\theta)=\psi^a(z)+\theta J^a(z)$ gives a
supercurrent associated with a Lie algebra, where $a$ is its index, in
the matter sector.\footnote{Its component fields satisfy following 
operator product expansions (OPE)
\cite{Mohammedi:1993rg}:
\begin{align}
&\psi^a(y)\psi^b(z)\sim \frac{1}{y-z}\frac{1}{2}\Omega^{ab},
~~~~J^a(y)\psi^b(z)\sim \frac{1}{y-z}f^{ab}_{~~\,c}\psi^c(z),
\nonumber\\
&J^a(y)J^b(z)\sim
  \frac{1}{(y-z)^2}\frac{1}{2}\Omega^{ab}+\frac{1}{y-z}f^{ab}_{~~\,c}J^c(z),
\nonumber
\end{align}
where constants $\Omega^{ab},f^{ab}_{~~\,c}$ satisfy following relations
\begin{align}
&\Omega^{ab}=\Omega^{ba},
~~~~~~f^{ab}_{~~\,c}\Omega^{cd}+f^{ad}_{~~\,c}\Omega^{cb}=0,
\nonumber\\
&f^{ab}_{~~\,c}=-f^{ba}_{~~\,c},
~~~~f^{ab}_{~~\,d}f^{cd}_{~~\,e}
+f^{bc}_{~~\,d}f^{ad}_{~~\,e}+f^{ca}_{~~\,d}f^{bd}_{~~\,e}=0.
\nonumber
\end{align}
}

The solution $\Phi_J$, which is in the large Hilbert space,
can be related to $\Psi_J$ which is regarded as a marginal
solution in the modified cubic superstring field theory \cite{Preitschopf:1989fc, Arefeva:1989cp}:
\begin{align}
&\Psi_J\equiv -V^a_L(F_a)I
+\frac{1}{8}\Omega^{ab}C_L(F_aF_b)I
=e^{-\Phi_J}Q_{\rm B}e^{\Phi_J},
\label{PsiJPhiJ}
\\
&V^a_L(f)\equiv\int_{C_{\rm L}}\frac{dz}{2\pi
 i}\frac{1}{\sqrt{2}}f(z)(cJ^a(z)+\gamma \psi^a(z)),
~~~C_L(f)\equiv\int_{C_{\rm L}}\frac{dz}{2\pi i}f(z)c(z).
\end{align}
{}From the expression in (\ref{PsiJPhiJ}), it can be easily found that
$\Phi_J$ satisfies the equation of motion in the NS sector,
\begin{align}
 \eta_0(e^{-\Phi_J}Q_{\rm B}e^{\Phi_J})=0,
\end{align}
because $\Psi_J$ is in the small Hilbert space.

By expanding the NS action $S[\Phi;Q_{\rm B}]$ of Berkovits'
WZW-like superstring field theory around $\Phi_J$ as
\begin{align}
 e^{\Phi}&=e^{\Phi_J}e^{\Phi'},
\end{align}
we have $S[\Phi;Q_{\rm B}]=S[\Phi_J;Q_{\rm B}]+S[\Phi';Q_{\Phi_J}]$,
where $S[\Phi';Q_{\Phi_J}]$ has the same form as the original action
except the BRST operator $Q_{\Phi_J}$:
\begin{align}
 Q_{\Phi_J}&=Q_{\rm B}-V^a(F_a)+\frac{1}{8}\Omega^{ab}C(F_aF_b).
\end{align}
Here, $V^a(F_a)$ and $C(F_aF_b)$ are given by integrations
along the whole unit circle, $|z|=1$:
\begin{align}
&V^a(f)\equiv\oint\frac{dz}{2\pi
 i}\frac{1}{\sqrt{2}}f(z)(cJ^a(z)+\gamma \psi^a(z)),
&C(f)\equiv\oint\frac{dz}{2\pi i}f(z)c(z).
\end{align}

\subsection{Deformed algebra}

Let us consider a version of the extended $KBc$ algebra in
\cite{Erler:2013wda}, in which the BRST operator is replaced with
a deformed BRST operator.
First, we give
the Virasoro generator, $L'_n$ corresponding to $Q'\equiv Q_{\Phi_J}$ 
\cite{Kishimoto:2005bs}:
\begin{align}
 L'_n&=\{Q',b_n\}=L_n
-\frac{1}{\sqrt{2}}\sum_{k\in {\mathbb{Z}}}F_{a,k}J^a_{n-k}
+\frac{1}{8}\Omega^{ab}\sum_{k\in{\mathbb{Z}}}F_{a,n-k}F_{b,k},
\end{align}
where we define the coefficients  as $F_{a,n}\equiv \oint
 \frac{d\sigma}{2\pi}e^{i(n+1)\sigma}F_a(e^{i\sigma})$
and then the condition $F_a(-1/z)=z^2F_a(z)$ imposes $F_{a,n}=-(-1)^nF_{a,-n}$.
We note that only the matter sector is deformed and the
 central charge is not changed.
Using this $L'_n$ instead of the conventional Virasoro generators, we define 
a string field $K'$ as in \cite{Inatomi:2012nd}.
Then, among string fields such as $K',B,c,\gamma$, we have the following relations:\footnote{These string
fields are given by
\begin{align}
&
B=\frac{\pi}{2}B_1^LI\sigma_3,~~~
c=\frac{2}{\pi}\hat{U}_1\tilde c(0)|0\rangle\sigma_3,
~~~
\gamma=\sqrt{\frac{2}{\pi}}\hat{U}_1\tilde \gamma(0)|0\rangle\sigma_2
~~~
K'=\frac{\pi}{2}K_1^{\prime L}I,~~~K=\frac{\pi}{2}K_1^LI,
\nonumber
\end{align}
where $K_1^{\prime L}\equiv \{Q',B_1^L\}$.
We use the conventions in \cite{Schnabl:2005gv} for
the definitions of 
$K_1^L,B_1^L,\hat{U}_r\equiv U^{\dagger}_rU_r,U_r=(2/r)^{{\cal L}_0},\cdots$ and
$\tilde c(\tilde z),\tilde \gamma(\tilde z),\cdots$ denote worldsheet fields in the sliver frame.
}
\begin{align}
&B^2=0,~~~c^2=0,~~~Bc+cB=1,~~~
BK'=K'B,~~~K'c-cK'=Kc-cK\equiv \partial c,\nonumber\\
&\gamma B+B\gamma=0,~~~~c\gamma+\gamma c=0,
~~~~K'\gamma-\gamma K'=K\gamma-\gamma K
\equiv\partial \gamma,\nonumber\\
&\hat{Q}'B=K',~~~\hat{Q}'K'=0,~~~\hat{Q}'c=cK'c-\gamma^2=cKc-\gamma^2
=c\partial c -\gamma^2,\nonumber\\
&\hat Q'\gamma=\hat Q\gamma=c\partial \gamma-\frac{1}{2}(\partial
  c)\gamma,
\end{align}
where $\hat{Q}'\equiv Q'\sigma_3,\hat{Q}\equiv Q_{\rm B}\sigma_3$
and $\sigma_i$ ($i=1,2,3$) are Pauli matrices attached as
the Chan-Paton factor in order to take into account
the GSO$(-)$ sector.

Furthermore, for the string fields $\gamma^{-1}, \zeta$ and 
$V$, we see more relations:\footnote{As in \cite{Erler:2013wda}, they are defined by
\begin{align}
&\gamma^{-1}=\sqrt{\frac{\pi}{2}}\hat{U}_1\tilde
 \gamma^{-1}(0)|0\rangle\sigma_2,
~~~\zeta=\gamma^{-1}c=\sqrt{\frac{2}{\pi}}\hat{U}_1\tilde\gamma^{-1}\tilde
 c(0)|0\rangle i\sigma_1,\ \ \ 
V=\frac{1}{2}\gamma^{-1}\partial c=
\sqrt{\frac{\pi}{2}}\hat{U}_1\frac{1}{2}\tilde\gamma^{-1}\tilde\partial
 \tilde c(0)|0\rangle i\sigma_1.
\nonumber
\end{align}
}
\begin{align}
&\gamma^{-1}\gamma=\gamma\gamma^{-1}=1,~~~
\gamma^{-1}B+B\gamma^{-1}=0,
~~~
\gamma^{-1}c+c\gamma^{-1}=0,\nonumber\\
&K'\gamma^{-1}-\gamma^{-1}K'=K\gamma^{-1}-\gamma^{-1}K\equiv
 \partial\gamma^{-1},\nonumber\\
&\hat{Q}'\gamma^{-1}=\hat{Q}\gamma^{-1}=c\partial \gamma^{-1}
+\frac{1}{2}(\partial c)\gamma^{-1},
~~~~~~
\hat{Q}'\zeta =\hat{Q}\zeta=cV+\gamma.
\end{align}
It is convenient to note further relations among string fields:
\begin{align}
&B\hat{Q}'\zeta+(\hat{Q}'\zeta)B=V,
~~~~B\hat{Q}'c-(\hat{Q}'c)B=\partial c,
~~~~\zeta^2=0,
~~~~c\zeta=\zeta c=0,\nonumber\\
&\gamma\zeta=-\zeta\gamma=c,
~~~~(\hat{Q}'\zeta)\zeta=-\zeta\hat{Q}'\zeta=c,
~~~~(\hat{Q}'c)\zeta=\zeta\hat{Q}'c=-\gamma c,\nonumber\\
&(\hat{Q}'\zeta)c=-c\hat{Q}'\zeta=\gamma c,
~~~~(\hat{Q}'\zeta)^2=-\hat{Q}'c,
\end{align}
for concrete computations.

It is noted that $K'$, $B$, $c$, $\gamma$, $\gamma^{-1}$, $\zeta$, $V$ and $\hat{Q}'$
have the same algebraic structure as that of the extended $KBc$ algebra
\cite{Erler:2013wda} with $\hat{Q}$.

\subsection{Tachyon vacuum solution
\label{sec:Tachyonvacuum}
}

From the result in \cite{Erler:2013wda} and the deformed algebra in the
previous subsection,
we can immediately construct a solution $\Phi_T$ in the theory with
$\hat{Q}'$:
\begin{align}
e^{\Phi_T}&=(1+q\zeta)\left(1+(q^2-1)c\frac{B}{1+K'}
+q(\hat{Q}'\zeta)\frac{B}{1+K'}\right)
\nonumber\\
&=1-c\frac{B}{1+K'}+q\left(\zeta+(\hat{Q}'\zeta)\frac{B}{1+K'}\right),
\label{ePhi_Tdef}
\end{align}
where $q$ is a nonzero constant. This solution has the form
in which $K$ of the Erler solution is replaced by $K'$.
In the same way as \cite{Erler:2013wda}, we find that 
$\Phi_T$ satisfies 
\begin{align}
e^{-\Phi_T}\hat{Q}'e^{\Phi_T}&=
c-(\hat{Q}'c)\frac{B}{1+K'}=(c+\hat{Q}'(Bc))\frac{1}{1+K'},
\label{e-PhiTQ'ePhiT}
\end{align}
which is in the small Hilbert space,
and therefore the equation of motion in the NS sector:
$\hat{\eta}(e^{-\Phi_T}\hat{Q}'e^{\Phi_T})=0$,
where $\hat{\eta}\equiv \eta_0\sigma_3$,
holds.

If we expand the action $S[\Phi';\hat{Q}']$ around the solution
$\Phi_T$ as $e^{\Phi'}=e^{\Phi_T}e^{\Phi''}$, 
we have a new BRST operator $\hat{Q}'_{\Phi_T}$ as
\begin{align}
 \hat{Q}'_{\Phi_T}\,\Xi
&=\hat{Q}'\,\Xi+(e^{-\Phi_T}\hat{Q}'e^{\Phi_T})\Xi -(-1)^{|\Xi|}\Xi(e^{-\Phi_T}\hat{Q}'e^{\Phi_T}),
\label{Q'PhiTdef}
\end{align}
where $\Xi$ is an arbitrary string field.
$|\Xi|$ denotes effective Grassmann parity of $\Xi$, which is defined by
a sum of Grassmann parity and worldsheet
spinor number of the vertex operator corresponding to $\Xi$ \cite{Erler:2010pr, Erler:2013wda}.
We note that the string field $e^{-\Phi_T}\hat{Q}'e^{\Phi_T}$  (\ref{e-PhiTQ'ePhiT})
 is essentially the same as the tachyon vacuum solution
on the marginally deformed background in the modified cubic superstring field theory
\cite{Inatomi:2012nd}.
Hence, we can find a homotopy operator $\hat{A}'$ for
$\hat{Q}'_{\Phi_T}$: 
\begin{align}
&\hat{A}'\,\Xi = \frac{1}{2}(A'\,\Xi+(-1)^{|\Xi|}\Xi\,A'),
\end{align}
for any string field $\Xi$, where $A'\equiv \frac{B}{1+K'}$ is a homotopy state in the small Hilbert space.
Actually, we find $\hat{Q}'_{\Phi_T}A'=1$ and $(A')^2=0$, which imply the following relations:
\begin{align}
&\{\hat{Q}'_{\Phi_T},\hat{A}'\}=1,
&(\hat{A}')^2=0.
\end{align}

Accordingly, the existence of the homotopy operator $\hat{A}'$ indicates that 
there is no physical open string state around the solution $\Phi_T$ and
therefore it is the tachyon vacuum solution in the
marginally deformed background.

\subsection{Energy and gauge invariant overlaps
\label{sec:Energy-GIO}
}

The NS action $S[\Phi';\hat{Q}_{\Phi_J}]$ around $\Phi_J$ is given by
\begin{align}
S[\Phi';\hat{Q}_{\Phi_J}]&=-\int_0^1dt \,{\rm
 Tr}\left[\left(\hat{\eta}(g(t)^{-1}\partial_tg(t))\right)
(g(t)^{-1}\hat{Q}_{\Phi_J} g(t))\right].
\end{align}
Here we have used a formula in \cite{Berkovits:2004xh} instead of the conventional
WZW-like form and
$g(t)$ ($t\in [0,1]$)
in the action is an interpolating string field  such that
$g(0)=1$ and $g(1)=e^{\Phi'}$.
The trace in the action implies ${\rm Tr}\, A\equiv \frac{1}{2}{\rm
tr}\langle I|A\rangle$, where ${\rm tr}$ is the trace for 
$2\times 2$ matrices corresponding to the Chan-Paton factor. 

In order to evaluate the action for $\Phi'=\Phi_T$, which is given in (\ref{ePhi_Tdef}), we take 
an interpolating string field:
\begin{align}
 g_T(t)&=1+t(e^{\Phi_T}-1)\nonumber\\
&=1-tc\frac{B}{1+K'}+qt\left(\zeta+(\hat{Q}'\zeta)\frac{B}{1+K'}\right),
\label{gt_def}
\end{align}
after \cite{Erler:2013wda}. 
The integrand in the action for the solution: $S[\Phi_T;\hat{Q}_{\Phi_J}]$, can be manipulated
in the same manner as the Erler solution. 
(Hence, we describe only an outline of our computations in the following.)
Using graded cyclicity of the trace and
${\rm Tr}[\hat{\eta}A]=0, {\rm Tr}[\hat{Q}'A]=0$
for any string field $A$, we find
\begin{align}
&{\rm Tr}\left[\left(\hat{\eta}(g_T(t)^{-1}\partial_tg_T(t))\right)
(g_T(t)^{-1}\hat{Q}_{\Phi_J} g_T(t))\right]
\nonumber\\
&=q(1-t){\rm Tr}\left[
(\hat{\eta}\hat{Q}'\zeta)
\left(
\frac{2q^2t^2-t}{{\cal D}'}Bc\frac{1}{{\cal D}'}
+\frac{qt}{{\cal D}'}B(\hat{Q}'\zeta)\frac{1}{{\cal D}'}
\right)
\right],
\end{align}
where ${\cal D}'\equiv 1-t+q^2t^2+K'+qt V$.
Because the trace becomes zero except for terms with $bc$-ghost number
$3$, we expand $\frac{1}{{\cal D}'}$ as
$\frac{1}{{\cal D}'}=\frac{1}{{\cal D}''}\sum_{n=0}^{\infty}
(-1)^n\left(qtV \frac{1}{{\cal D}''}\right)^n$
(${\cal D}''\equiv 1-t+q^2t^2+K'$)
and the above trace can be computed as
\begin{align}
&{\rm Tr}\left[\left(\hat{\eta}(g_T(t)^{-1}\partial_tg_T(t))\right)
(g_T(t)^{-1}\hat{Q}_{\Phi_J} g_T(t))\right]=
q^2t^2(1-t)(2q^2t-1){\cal X}_1+
q^2t(1-t){\cal X}_2,\\
&{\cal X}_1={\rm Tr}\left[
(\hat{\eta}{\hat Q}'\zeta)\frac{1}{{\cal D}''}
cV\frac{B}{({\cal D}'')^2}\right],
~~~{\cal X}_2={\rm Tr}\left[
(\hat{\eta}\hat{Q}'\zeta)B\frac{1}{{\cal D}''}cV\frac{1}{{\cal D}''}
\right].
\end{align}
In order to evaluate ${\cal X}_1$ and ${\cal X}_2$, 
we note the following equations, related to a derivation 
${\cal L}_0'-{\cal L}_0^{\prime \dagger}=\{Q',{\cal B}_0-{\cal B}_0^{\dagger}\}$ with respect to
the star product among string fields,
 as in \cite{Inatomi:2012nd, Erler:2013wda}. Namely,
\begin{align}
{\rm Tr}\left[({\cal L}_0'-{\cal L}_0^{\prime \dagger})A\right]=0,
\label{scale}
\end{align}
for any string field $A$ and 
\begin{align}
&\frac{1}{2}({\cal L}_0'-{\cal L}_0^{\prime \dagger})c=-c,
~~~~\frac{1}{2}({\cal L}_0'-{\cal L}_0^{\prime \dagger})B=B,
~~~~
\frac{1}{2}({\cal L}_0'-{\cal L}_0^{\prime \dagger})\gamma
=-\frac{1}{2}\gamma,
~~~~
\frac{1}{2}({\cal L}_0'-{\cal L}_0^{\prime \dagger})K'=K',
\nonumber\\
&\frac{1}{2}({\cal L}_0'-{\cal L}_0^{\prime \dagger})\gamma^{-1}
=\frac{1}{2}\gamma^{-1},
~~~~\frac{1}{2}({\cal L}_0'-{\cal L}_0^{\prime \dagger})V=
\frac{1}{2}V,
~~~~
\frac{1}{2}({\cal L}_0'-{\cal L}_0^{\prime \dagger})\zeta
=-\frac{1}{2}\zeta.
\end{align}
In particular, using
\begin{align}
 e^{\frac{1}{2}({\cal L}_0'-{\cal L}_0^{\prime \dagger})
\log(1-t+q^2t^2)}
\frac{1}{{\cal D}''}=(1-t+q^2t^2)^{-1}\frac{1}{1+K'},
\end{align}
the traces are simplified as
\begin{align}
{\cal X}_1&=-\frac{1}{(1-t+q^2t^2)^3}{\rm Tr}\left[
B(\hat{\eta}(cV))\frac{1}{1+K'}cV
\frac{1}{(1+K')^2}
\right],
\\
{\cal X}_2&=\frac{1}{(1-t+q^2t^2)^2}{\rm Tr}\left[
B(\hat{\eta}(cV))\frac{1}{1+K'}cV\frac{1}{1+K'}
\right].
\end{align}
In order to calculate these traces, we use an expression with the Schwinger parameter
for the inverse of $1+K'$: 
\begin{align}
 \frac{1}{1+K'}&=\int_0^{\infty} d\alpha\, e^{-\alpha}e^{-\alpha K'}.
\end{align}
Performing some change of variables in the integrations and 
inserting $e^{\frac{1}{2}({\cal L}_0'-{\cal L}_0^{\prime
\dagger})(-\log\ell)}$ in the trace appropriately, we reach the
following expressions: 
\begin{align}
 {\cal X}_1&=\frac{-2}{(1-t+q^2t^2)^3}
\int_0^1\!\!d\theta\,(1-\theta){\rm Tr}\!\left[
B(\hat{\eta}(cV))e^{-\theta K'}cV
e^{-(1-\theta)K'}
\right],
\label{cX1f}
\\
{\cal
 X}_2&=\frac{1}{(1-t+q^2t^2)^2}\int_0^1 d\theta\,
{\rm Tr}\left[
B(\hat{\eta}(cV))e^{-\theta K'}cVe^{-(1-\theta)K'}
\right].
\label{cX2f}
\end{align}
For both of them, we need to calculate
the trace ${\rm Tr}\!\left[
B(\hat{\eta}(cV))e^{-\theta K'}cV
e^{-(1-\theta)K'}
\right]$.
Unlike the case of \cite{Erler:2013wda}, $e^{-\alpha K'}$ appears in
the above trace instead of $e^{-\alpha K}$. Here,
we use the result for the modified cubic superstring field theory in the
marginally deformed background \cite{Inatomi:2012nd}:
\begin{align}
&e^{-\alpha K'}=e^{-\alpha\frac{\pi}{2}{\cal C}}\hat{U}_{\alpha+1}
{\bf T}\exp\!\left(
\frac{\pi}{4}\int_{-\alpha}^{\alpha}du\int_{-\infty}^{\infty}dv
f_a(v)\tilde J^a(iv+\frac{\pi}{4}u)
\right)\!|0\rangle,\\
&f_a(v)\equiv \frac{F_a(\tan(iv+\frac{\pi}{4}))}{2\pi\sqrt{2}
\cos^2(it+\frac{\pi}{4})},
~~~~~~~
{\cal C}\equiv
 \frac{\pi}{2}\int_{-\infty}^{\infty}dv\,\Omega^{ab}f_a(v)f_b(v),
\end{align}
where ${\bf T}$ is an ordering symbol
with respect to the real part of the argument of $\tilde J^a$.
The star products in the trace can be calculated as
\begin{align}
B(\hat{\eta}(cV))
 e^{-\theta K'}cV e^{-(1-\theta)K'}
&=
\frac{e^{-\frac{\pi}{2}{\cal C}}}{4}
B_1^L(\eta_0\tilde \gamma^{-1})\tilde
 c\tilde\partial\tilde c(\frac{\pi}{4})
\tilde \gamma^{-1}\tilde
 c\tilde\partial\tilde c(\frac{\pi}{4}(1-2\theta))
\nonumber\\
&~~~\times {\bf T}\exp\!\left(\!
\frac{\pi}{4}\!\int_{-1}^1\!\!\!du\!
\int_{-\infty}^{\infty}\!\!\!\!dv
f_a(v)\tilde J^a(iv+\!\frac{\pi}{4}u)\!
\right)\!|0\rangle.
\end{align}
Finally, the trace appeared in (\ref{cX1f}) and (\ref{cX2f}) can be
evaluated as
\begin{align}
&{\rm Tr}\!\left[
B(\hat{\eta}(cV))e^{-\theta K'}cV
e^{-(1-\theta)K'}\right]
\nonumber\\
&=-\frac{1}{4}\left\langle
(\eta_0\gamma^{-1}(\frac{\pi}{2}))\gamma^{-1}(\frac{\pi}{2}(1-2\theta))
\right\rangle_{\xi\eta\phi}
\left\langle
B_1^Lc\partial c(\frac{\pi}{2})c\partial c(\frac{\pi}{2}(1-2\theta))
\right\rangle_{bc}
\nonumber\\
&~~~~~~~~~~\times e^{-\frac{\pi}{2}{\cal C}}
\left\langle 
\exp\!\left(\int_{-\frac{\pi}{2}}^{\frac{\pi}{2}}\!\!\!du\!
\int_{-\infty}^{\infty}\!\!\!\!dv
f_a(v)J^a(2iv+u)\!
\right)
\right\rangle_{\rm mat}.
\end{align}
Here, the correlation functions are defined on a cylinder of
circumference $\pi$.
The last factor on the right hand side, which is a
correlator in  the matter sector, is $1$ as proved in
\cite{Inatomi:2012nd}, and so 
the trace becomes the
same result as the case of the Erler solution.
Consequently, the vacuum energy of (\ref{ePhi_Tdef}) is unchanged from
the case in the original background without marginal
deformations, namely, $E=-S[\Phi_T;\hat{Q}_{\Phi_J}]=-1/(2\pi^2)$. \\

Next, we evaluate the gauge invariant overlap (GIO)
in Berkovits' WZW-like superstring field theory.
We define the GIO $\langle\Phi\rangle_{\cal V}$ as
\begin{align}
\langle\Phi\rangle_{\cal V}&\equiv {\rm Tr}[{\cal V}(i)\Phi]
\label{GIOdef}
\end{align}
after \cite{Michishita:2004rx}. 
Here, ${\cal V}(i)$ denotes a midpoint insertion of a primary closed string vertex operator with 
the picture $-1$, the ghost number $2$ and the conformal dimension $(0,0)$
and it should be BRST invariant and in the small Hilbert space, i.e., 
$[Q_{\rm B},{\cal V}(i)]=0$ and $[\eta_0,{\cal V}(i)]=0$.
It satisfies
\begin{align}
&\langle\hat{Q}\Lambda\rangle_{\cal V}=0,
&&\langle\hat{\eta}\Lambda\rangle_{\cal V}=0,
&&\langle \Lambda\Xi\rangle_{\cal V}=(-)^{|\Lambda||\Xi|}\langle\Xi\Lambda\rangle_{\cal V}
\label{GIOprop}
\end{align}
for any string fields $\Lambda,\Xi$.
We note that an infinitesimal gauge transformation
\begin{align}
\delta_{\Lambda} e^{\Phi}&=(\hat{Q}\Lambda_0)e^{\Phi}+e^{\Phi} \hat{\eta}\Lambda_{1},
\end{align}
($\Lambda_0$ and $\Lambda_1$ are gauge parameter string field with the picture number 
$0$ and $1$, respectively.)
which can be rewritten as\footnote{ We have used
\begin{align}
 e^{-\Phi}\delta e^{\Phi}&=\int_0^1d\theta\,
 e^{-\theta\Phi}(\delta\Phi)e^{\theta\Phi}
=\int_0^1d\theta\, e^{-\theta{\rm ad}_{\Phi}}\delta\Phi
=\frac{e^{-{\rm ad}_{\Phi}}-1}{-{\rm ad}_{\Phi}}\delta\Phi.
\end{align}
}
\begin{align}
 \delta_{\Lambda}\Phi&=\frac{{\rm ad}_{\Phi}}{e^{{\rm
 ad}_{\Phi}}-1}\hat{Q}\Lambda_0
+\frac{-{\rm ad}_{\Phi}}{e^{-{\rm
 ad}_{\Phi}}-1}\hat{\eta}\Lambda_1,
\end{align}
where ${\rm ad}_B(A)\equiv [B,A]=BA-AB$.
Thanks to the properties (\ref{GIOprop}),
the GIO is invariant under the above transformation: $\langle \delta_{\Lambda}\Phi\rangle_{\cal V}=0$.
We note that the GIO for $\Phi'$ in the theory around a solution $\Phi_0$
is also invariant under the gauge transformation 
because $\hat{Q}\Lambda_0$ is replaced with 
 $\hat{Q}_{\Phi_0}\Lambda_0\equiv \hat{Q}\Lambda_0+[e^{-\Phi_0}\hat{Q}e^{\Phi_0},\Lambda_0]$
in the above.

Let us evaluate the GIO for $\Phi_T$.
It is convenient to use $e^{-\Phi_T}\hat{Q}'e^{\Phi_T}$ (\ref{e-PhiTQ'ePhiT}) 
instead of $\Phi_T$ itself.
Inserting $1=\{Q_{\rm B},\xi Y(i)\}$, 
where $Y(z)=c\partial \xi e^{-2\phi}(z)$ is the inverse picture changing operator,
the GIO (\ref{GIOdef}) can be rewritten as\footnote{
Similarly, it can be expressed as \cite{Erler:2013wda}:
\begin{align}
\langle\Phi\rangle_{\cal V}&={\rm Tr}[\xi Y{\cal V}(i)\sigma_3\,e^{-\Phi}\hat{Q}e^{\Phi}]
={\rm Tr}[\xi {\cal V}(i)\sigma_3\,e^{-\Phi}\hat{\eta}e^{\Phi}]
=\int_0^1dt\,{\rm Tr}[{\cal V}(i)\,e^{-\Phi(t)}\partial_t e^{\Phi(t)}],
\label{GIO3exp}
\end{align}
where $\Phi(t)$ is an interpolation such as $\Phi(0)=0,\Phi(1)=\Phi$.
}
\begin{align}
\langle\Phi\rangle_{\cal V}&={\rm Tr}[{\cal V}(i)\{Q_{\rm B},\xi Y(i)\}\Phi]
={\rm Tr}[\xi Y{\cal V}(i)\sigma_3\hat{Q}\Phi]
={\rm Tr}[\xi Y{\cal V}(i)\sigma_3\hat{Q}'\Phi]
\nonumber\\
&={\rm Tr}[\xi Y{\cal V}(i)\sigma_3\,e^{-\Phi}\hat{Q}'e^{\Phi}].
\label{t2Q'}
\end{align}
Substituting  (\ref{e-PhiTQ'ePhiT}), we have
\begin{align}
\langle\Phi_T\rangle_{\cal V}&={\rm Tr}\left[\xi Y{\cal V}(i)\sigma_3\, c\frac{1}{1+K'}\right].
\end{align}
In a similar way to the calculation of the vacuum energy, by using the
Schwinger representation and the result of the correlation function in
the modified cubic superstring field theory \cite{Inatomi:2012nd}, we finally obtain the expression
of the GIO in the marginally deformed background:\footnote{
Here, we have used an expression in \cite{Kishimoto:2013sra}
rather than that in \cite{Inatomi:2012nd}.
}
\begin{align}
\langle\Phi_T\rangle_{\cal V}&=\frac{e^{-\pi{\cal C}}}{\pi}
\left\langle
\xi Y{\cal V}(i\infty)c(\frac{\pi}{2})
\exp\left(\int_{-\frac{\pi}{2}}^{\frac{\pi}{2}}
\!\!\!du\,{\cal J}(u)\right)
\right\rangle_{C_\pi},
\label{tGIOPhi_Tf}
\\
{\cal J}(u)&=\int_{-\infty}^{\infty}dv\,f_a(v)J^a(iv+u),
\end{align}
where $\langle\cdots \rangle_{C_\pi}$ denotes the correlation functions
on a cylinder of circumference $\pi$.

\section{Analytic evaluation of observables for
  identity-based marginal solutions
\label{sec:analytic}
}

In this section, we consider the gauge equivalence relation mentioned in
the introduction and we calculate two observables, the vacuum
energy and the gauge invariant overlap, for the identity-based marginal
solutions. 

First of all, we take a particular interpolation 
$\Phi_J(t)=t\Phi_J$ ($t\in [0,1]$) such that $\Phi_J(0)=0$ and
$\Phi_J(1)=\Phi_J$. 
This interpolating string field $\Phi_J(t)$ is also an
identity-based marginal solution to the equation of motion: 
\begin{align}
 \hat{\eta}(e^{-\Phi_J(t)}\hat{Q}e^{\Phi_J(t)})=0,
\label{EOMhatPhiJ}
\end{align}
because $\Phi_J(t)$ is given by a replacement of 
the weighting function: $F_a(z)\to tF_a(z)$ in (\ref{marginalsol}).
Hence, we have a new BRST operator $Q_{\Phi_J(t)}$ :
\begin{align}
 Q_{\Phi_J(t)}&=Q_{\rm B}-tV^a(F_a)+\frac{t^2}{8}\Omega^{ab}C(F_aF_b),
\end{align}
for the theory around a solution $\Phi_J(t)$.
Following the same procedure in \S\ref{sec:K'Bc}, we define a string
field $K'(t)\equiv \hat{Q}_{\Phi_J(t)}B$ ($\hat{Q}_{\Phi_J(t)}\equiv
Q_{\Phi_J(t)}\sigma_3$). Then we can construct
a tachyon vacuum solution $\Phi_T(t)$ as
\begin{align}
 e^{\Phi_T(t)}&=1-c\frac{B}{1+K'(t)}+q\left(\zeta+(\hat{Q}_{\Phi_J(t)}\zeta)
\frac{B}{1+K'(t)}\right),
\end{align}
and we can easily see that it satisfies the equation of motion around
the identity-based solution $\Phi_J(t)$:
\begin{align}
 \hat{\eta}(e^{-\Phi_T(t)}\hat{Q}_{\Phi_J(t)}e^{\Phi_T(t)})=0.
\label{EOMTt}
\end{align}
We note that, in particular, $\Phi_T(t)$ satisfies $\Phi_T(1)=\Phi_T$
(the solution (\ref{ePhi_Tdef}))
and $\Phi_T(0)=\Phi_T^{\rm E}$ (the Erler solution \cite{Erler:2013wda})
because $Q_{\Phi_J(1)}=Q_{\Phi_J}$ and $Q_{\Phi_J(0)}=Q_{\rm B}$.

Using the above string fields, we define a string field $\tilde\Phi_T(t)$
with the parameter $t$ as
\begin{align}
 e^{\tilde \Phi_T(t)}&\equiv e^{\Phi_J(t)}e^{\Phi_T(t)},
\end{align}
and then we find a relation:
\begin{align}
e^{-\tilde{\Phi}_T(t)}\hat{Q}e^{\tilde{\Phi}_T(t)}
&= e^{-\Phi_J(t)}\hat{Q}e^{\Phi_J(t)}+
e^{-\Phi_T(t)}\hat{Q}_{\Phi_J(t)}e^{\Phi_T(t)}.
\label{rele-Qe}
\end{align}
It indicates that, with (\ref{EOMhatPhiJ}) and (\ref{EOMTt}), $\tilde\Phi_T(t)$
satisfies the equation of motion of the original theory:
\begin{align}
 \hat{\eta}(e^{-\tilde\Phi_T(t)}\hat{Q}e^{\tilde\Phi_T(t)})=0.
\end{align}
Expanding around the solution $\tilde\Phi_T(t)$ in the theory with $\hat{Q}$,
we have the theory with the deformed BRST operator
$\hat{Q}_{\tilde{\Phi}_T(t)}$ such that, for any string field $\Xi$,
\begin{align}
 \hat{Q}_{\tilde{\Phi}_T(t)}\Xi
&=\hat{Q}\Xi+(e^{-\tilde\Phi_T(t)}\hat{Q}e^{\tilde\Phi_T(t)})\Xi 
-(-1)^{|\Xi|}\Xi(e^{-\tilde\Phi_T(t)}\hat{Q}e^{\tilde\Phi_T(t)})\nonumber\\
&=\hat{Q}_{\Phi_J(t)}\Xi+(e^{-\Phi_T(t)}\hat{Q}_{\Phi_J(t)}e^{\Phi_T(t)})\Xi
-(-1)^{|\Xi|}\Xi (e^{-\Phi_T(t)}\hat{Q}_{\Phi_J(t)}e^{\Phi_T(t)}).
\end{align}
Comparing to (\ref{Q'PhiTdef}), the last expression in the above implies that $\hat{Q}_{\tilde{\Phi}_T(t)}$ is the same as
the BRST operator $\hat{Q}'_{\Phi_T(t)}$ 
in the theory around $\Phi_T(t)$, which is a tachyon vacuum solution 
in the theory around $\Phi_J(t)$. 
Following the previous results in \S \ref{sec:Tachyonvacuum}
with appropriate replacement, we find that there exists a homotopy state: $A'(t)\equiv\frac{B}{1+K'(t)}$
such as $\hat{Q}_{\tilde{\Phi}_T(t)}A'(t)=1$,
which implies that there is no cohomology for $\hat{Q}_{\tilde{\Phi}_T(t)}$ in the small Hilbert space.
We also find the equation:\footnote{
In general, $\Psi(t)\equiv e^{-\Phi(t)}\hat{Q}e^{\Phi(t)}$ satisfies
$\hat{Q}\Psi(t)+\Psi(t)^2=0$. Differentiating it with respect to $t$, we find 
\begin{align}
\hat{Q}_{\Phi(t)}\frac{d}{dt}\left(e^{-\Phi(t)}\hat{Q}e^{\Phi(t)}\right)=0.
\end{align}
}
\begin{align}
&\hat{Q}_{\tilde{\Phi}_T(t)}\frac{d}{dt}(e^{-\tilde\Phi_T(t)}
\hat{Q}e^{\tilde\Phi_T(t)})=0.
\end{align}
Therefore,
 there exists a state $\Lambda_t$ in the small Hilbert space such as
\begin{align}
&\frac{d}{dt}(e^{-\tilde\Phi_T(t)}\hat{Q}e^{\tilde\Phi_T(t)})
=\hat{Q}_{\tilde{\Phi}_T(t)}\Lambda_t.
\label{ddtPsit}
\end{align} 
Integrating (\ref{ddtPsit}), we have
\begin{align}
e^{-\tilde{\Phi}_T(1)}\hat{Q}e^{\tilde{\Phi}_T(1)}
&=e^{-\Phi_T^{\rm E}}\hat{Q}e^{\Phi_T^{\rm E}}
+\int_0^1\hat{Q}_{\tilde{\Phi}_T(t)}\Lambda_t\,dt.
\label{ge4}
\end{align}
This is a gauge equivalence relation 
between the Erler solution $\Phi_T^{\rm E}$ and
$\tilde{\Phi}_T(1)=\log(e^{\Phi_J}e^{\Phi_T})$.
(See Appendix A for details of gauge equivalence. This relation
corresponds to (\ref{ge3}).)

Now that the gauge equivalence is established, we can analytically
evaluate physical observables for the identity-based marginal solutions.
First, the vacuum energy for the solution is calculated by using
(\ref{veid}). 
We have seen in \S \ref{sec:Energy-GIO} that $S[\Phi_T;\hat{Q}_{\Phi_J}]$ is
equal to $S[\Phi_T^{\rm E};\hat{Q}]$ and then we conclude that the vacuum
energy of the identity-based marginal solution $\Phi_J$ is zero from (\ref{veid}):
$E=-S[\Phi_J;\hat{Q}]=0$. This result agrees with the previous one
\cite{Kishimoto:2005bs,Kishimoto:2005wv} derived from zero mode counting of $\xi$. \\

Next, let us consider the GIO for the identity-based marginal solution: $\langle\Phi_J\rangle_{\cal V}$.
Substituting the relation (\ref{rele-Qe}) with $t=1$ into (\ref{ge4}), 
we obtain the relation (\ref{idge}) mentioned in the introduction. 
Using the expressions for the GIOs, (\ref{GIO3exp}) and (\ref{t2Q'}),
it immediately leads to a relation for the GIOs: 
\begin{align}
\langle\Phi_J\rangle_{\cal V}&=\langle\Phi_T^{\rm E}\rangle_{\cal V}-\langle\Phi_T\rangle_{\cal V},\label{GIOPhi_Jdiff}
\end{align}
where the contribution of the last term in (\ref{idge}) vanishes  thanks to (\ref{GIOprop}).
More explicitly, using the formula (\ref{tGIOPhi_Tf}), the GIO for $\Phi_J$ can be
represented by a correlation function:
\begin{align}
\langle\Phi_J\rangle_{\cal V}&=\frac{1}{\pi}
\left\langle
\xi Y{\cal V}(i\infty)c(\frac{\pi}{2})
\Bigl\{
1-
e^{-\pi{\cal C}}\exp\Bigl(
\int_{-\frac{\pi}{2}}^{\frac{\pi}{2}} du\,{\cal J}(u)
\Bigr)
\Bigr\}
\right\rangle_{C_\pi}.
\label{GIOPhi_Jdiff2}
\end{align}
As in \cite{Kishimoto:2013sra}, it can be rewritten as
a difference between two disk amplitudes with the boundary deformation
by taking
\begin{align}
F_a(z;s)=\frac{2\lambda_a s(1-s^2)}{\arctan\frac{2s}{1-s^2}}
\frac{1+z^{-2}}{1-s^2(z^2+z^{-2})+s^4},
\end{align}
for the function $F_a(z)$ in $\Phi_J$,
which satisfies
\begin{align}
\int_{C_{\rm L}}\frac{dz}{2\pi i}F_a(z;s)&=\frac{2\lambda_a}{\pi},
\\
F_a(z;s)&\to 4\lambda_a\{\delta(\theta)+\delta(\pi-\theta)\},
~~~~~(s\to 1,~~z=e^{i\theta}).
\end{align}
Namely for the limit $s\to 1$, the GIO becomes\footnote{
In this expression, ${\cal C}$ becomes divergent
for the function $\lim_{s\to 1}F_a(z;s)$ and then it cancels the contact
term divergence due to singular OPE among the currents.
}
\begin{align}
\langle\Phi_J\rangle_{\cal V}&=\frac{1}{\pi}
\left\langle
\xi Y{\cal V}(i\infty)c(\frac{\pi}{2})
\Bigl\{
1-
e^{-\pi{\cal C}}\exp\Bigl(
\frac{\sqrt{2}}{\pi}\lambda_a
\int_{-\frac{\pi}{2}}^{\frac{\pi}{2}} du\,J^a(u)
\Bigr)
\Bigr\}
\right\rangle_{C_\pi}.
\label{GIOPhi_Jdiff3}
\end{align}
This expression corresponds to the result in \cite{Ellwood:2008jh} 
for a wedge-based marginal solution \cite{Kiermaier:2007ki}.

\section{Concluding remarks
\label{sec:Rem}}

In this paper, we have applied the method in \cite{Inatomi:2012nv, Inatomi:2012nd} 
to the Erler solution $\Phi_T^{\rm E}$ \cite{Erler:2013wda}.
We have constructed a tachyon vacuum solution $\Phi_T$
around the identity-based marginal solution $\Phi_J$ \cite{Kishimoto:2005bs} in the framework of Berkovits' 
WZW-like superstring field theory.
To get the solution, we have used an extended $KBc$ algebra 
in the marginally deformed background.
Around the solution, we have obtained a homotopy operator
and we have evaluated the vacuum energy and the gauge invariant overlap
(GIO) for it.
The energy is the same value as that on the undeformed background
but the GIO is deformed by the marginal operators.

Using the above, we have extended our computation 
for bosonic string field theory \cite{Kishimoto:2013sra} to superstring.
Then, we have evaluated the energy and the GIO for 
the identity-based marginal solution $\Phi_J$.
The energy for $\Phi_J$ vanishes and it is consistent with our previous
result in \cite{Kishimoto:2005bs}.
The GIO for $\Phi_J$ is expressed by a difference of those of
the tachyon vacuum solutions on the
undeformed and deformed backgrounds.
In the evaluation of the gauge invariants, the relation (\ref{idge}) is
essential, which implies the gauge equivalence of $\Phi_T^{\rm E}$ and
$\log(e^{\Phi_J}e^{\Phi_T})$.

It should be pointed out that the general arguments of the gauge
equivalence in Appendix \ref{sec:ge} is applicable to the bosonic open
string field theory,
where $\Psi$ is regarded as the bosonic string field 
and the relations (\ref{ge2}) and (\ref{ge3}), without the
$\eta_0$-constraint, are equivalent. 
Therefore, the calculation of the gauge invariant overlaps for the
identity-based marginal solution in \cite{Kishimoto:2013sra},
where a version of  (\ref{ge3}) was used,
 implies that the vacuum energy for it is also evaluated analytically to
 be zero from gauge equivalence.

Similarly, in the framework of the modified cubic superstring field
theory, we can evaluate the energy and the GIO for
the identity-based marginal solution
$\Psi_J=e^{-\Phi_J}\hat{Q}e^{\Phi_J}$ with the result in \cite{Inatomi:2012nd}. From the gauge equivalence
relations for the tachyon vacuum solution
in the original theory \cite{Erler:2007xt, Gorbachev:2010zz} and that in the marginally deformed background \cite{Inatomi:2012nd}, which we denote as $\Psi_T$ and $\Psi_T'$, respectively,
the energy of $\Psi_J$ vanishes because of 
$S[\Psi_J;\hat{Q}]=S[\Psi_T;\hat{Q}]-S[\Psi_T';\hat{Q}']=1/(2\pi^2)-1/(2\pi^2)=0$.
The GIO for $\Psi_J$, where
$\langle\!\langle \Psi_J\rangle\!\rangle_{\cal V}\equiv {\rm Tr}[{\cal V}(i)\sigma_3\Psi_J]$,
 is computed by the relation
$\langle\!\langle \Psi_J\rangle\!\rangle_{\cal V}
=\langle\!\langle \Psi_T\rangle\!\rangle_{\cal V}
-\langle\!\langle \Psi_T'\rangle\!\rangle_{\cal V}
$.
It is given by
replacing $\xi Y{\cal V}$ with ${\cal V}$ in (\ref{GIOPhi_Jdiff2}),
which is evaluated using the correlator in the small Hilbert space.
One might think that the value of the GIO for $\Psi_J$ may become one half 
if the half brane solutions, $\Psi_H$ and $\Psi_H'$, in the original theory \cite{Erler:2010pr} and 
the marginally deformed background \cite{Inatomi:2012nd}, respectively,
are used instead of the tachyon vacuum solutions due to the relations,
$\langle\!\langle \Psi_H\rangle\!\rangle_{\cal V}=\frac{1}{2}\langle\!\langle \Psi_T\rangle\!\rangle_{\cal V}$ and  $\langle\!\langle \Psi_H'\rangle\!\rangle_{\cal V}=\frac{1}{2}\langle\!\langle \Psi_T'\rangle\!\rangle_{\cal V}$.
However, in the case of the half brane solutions, we expect that the BRST operators around them
have non-trivial cohomology in contrast to vanishing cohomology around the tachyon vacuum.
It implies that we cannot derive the equation corresponding to (\ref{ddtPsit})\footnote{
In other words,  the solutions, $\Psi_T$ and $\Psi_J+\Psi_T'$, in the original theory 
are gauge equivalent, 
but the half brane solution $\Psi_H$ seems to be gauge inequivalent to 
the solution $\Psi_J+\Psi_H'$.
}
 and
therefore we cannot use
$\langle\!\langle \Psi_H\rangle\!\rangle_{\cal V}
-\langle\!\langle \Psi_H'\rangle\!\rangle_{\cal V}
$
to evaluate $\langle\!\langle \Psi_J\rangle\!\rangle_{\cal V}$.

We comment on an evaluation of
the GIO for the identity-based marginal solution $\Phi_J$
in the case that the current is given by the 9-th direction:
${\bm J}(z,\theta)=\psi^9(z)+\theta\frac{i}{\sqrt{2\alpha'}}\partial X^9(z)$.
For the operator $\tilde V_L(F)$ in (\ref{marginalsol}),
we have a relation
$\tilde V_L(F)=\{Q_{\rm B},\Omega_L(F)\}+X_L(F)$,
where
$\Omega_L(F)\equiv \int_{C_{\rm L}}\frac{dz}{2\pi i}F(z)\frac{i}{2\sqrt{\alpha'}}\xi Y X^9(z)$
and $X_L\equiv \int_{C_{\rm L}}\frac{dz}{2\pi i}F(z)\frac{-i}{2\sqrt{\alpha'}}X^9(z)$.
Therefore, the GIO might be computed such that
$\langle \Phi_J\rangle_{\cal V}=\langle \tilde V_L(F)I\rangle_{\cal V}=\langle X_L(F)I\rangle_{\cal V}$,
which formally vanishes\footnote{
This argument for vanishing GIO corresponds to that in \cite{Katsumata:2004cc}
for the bosonic case.
}
due to the lack of the $\xi$ zero mode.
However, it is not consistent with our result (\ref{GIOPhi_Jdiff2}),
which does not vanishes for $Y{\cal
V}(i\infty)=ce^{-\phi}e^{\frac{i}{\sqrt{2\alpha'}}X^9}(i\infty)\,ce^{-\phi}e^{-\frac{i}{\sqrt{2\alpha'}}X^9}(-i\infty)$
as evaluated in \cite{Inatomi:2012nd}.
More explicitly, we have
$\langle \Phi_J\rangle_{\cal V}=\frac{1}{2\pi i}(1-e^{i\pi \sqrt{2}f})$,
where $f\equiv \int_{C_{\rm L}}\frac{dz}{2\pi i}F(z)$.
This apparent discrepancy seems to imply that $\langle X_L(F)I\rangle$ is 
ill-defined because of a singular property of the trace of
identity-based states.
In the case of $f=0$, the solution
$\Phi_J$ becomes pure gauge \cite{Kishimoto:2005bs} and it is consistent with $\langle \Phi_J\rangle_{\cal V}=0$.

\section*{Acknowledgements}
We would like to thank Toru Masuda for helpful comments.
The work of I.~K. and T.~T. is supported by JSPS Grant-in-Aid for
Scientific Research (B) (\#24340051). 
The work of I.~K. is supported in part by
JSPS Grant-in-Aid for Young Scientists (B) (\#25800134).

\appendix

\section{On gauge equivalence relations
\label{sec:ge}
}

Here, we discuss some gauge equivalence relations in terms of the NS sector of
Berkovits' WZW-like superstring field theory.

Let us consider a gauge transformation of the superstring field
$\Phi$ by group elements $h(t)$ and $g(t)$ with one parameter $t$
such that $g(0)=h(0)=1$:
\begin{align}
&e^{\Phi(t)}= h(t)\,e^\Phi\,g(t),
&Q_{\rm B} h(t)=\eta_0 g(t)=0.
\label{ge1}
\end{align}
For the string fields, $\Phi$ and $\Phi(t)$, ``one-form'' string fields are
defined by $\Psi\equiv e^{-\Phi}\,Q_{\rm B} e^\Phi$ and
$\Psi(t)\equiv e^{-\Phi(t)}\,Q_{\rm B}e^{\Phi(t)}$. From (\ref{ge1}), these turn out to be related by
a transformation:
\begin{align}
&\Psi(t)=g(t)^{-1}\,Q_{\rm B}\,g(t)+g(t)^{-1}\Psi\,g(t),
&\eta_0g(t)=0.
\label{ge2}
\end{align}
It is the same form with the gauge transformation in the NS sector of the modified cubic 
superstring field theory.
Conversely, given the relation (\ref{ge2}), we
find that the relation (\ref{ge1}) holds for 
$h(t)=e^{\Phi(t)} g(t)^{-1} e^{-\Phi}$.
In fact, from (\ref{ge2}), we immediately see that $Q_{\rm B}(e^{\Phi(t)} g(t)
e^{-\Phi})=0$ holds.

Differentiating (\ref{ge2}) with respect to $t$ and integrating it again,
we find another relation between $\Psi$ and $\Psi(t)$:
\begin{align}
&\Psi(t)=\Psi+\int_0^t
 \,Q_{\Phi(t')}\Lambda(t')\,dt',
&&\eta_0\Lambda(t)=0,
\label{ge3}
\end{align}
where $\Lambda(t)=g(t)^{-1}\,\frac{d}{dt}g(t)$ and
$Q_{\phi}$ is a modified BRST operator associated with 
$\psi\equiv e^{-\phi}\,Q_{\rm B} e^{\phi}$:
$Q_{\phi}\lambda=Q_{\rm B}\lambda+\psi\lambda-(-1)^{|\lambda|}\psi \lambda$
for any string field $\lambda$.
Conversely, supposing that the equations (\ref{ge3}) for a given $\Lambda(t)$ hold,
we find the relations ($\ref{ge2}$) hold for the group element $g(t)$ such as $g(0)=1$:
\begin{align}
 g(t)&={\rm P}\exp \left(\int_0^t \Lambda(t')dt'\right),
\end{align}
where ${\rm P}\exp$ means a $t$-ordered exponent.

Consequently, the above relations (\ref{ge1}), (\ref{ge2}) and (\ref{ge3}) are
all equivalent. It is noted that we can easily include internal Chan-Paton
factors in these relations.


\end{document}